# MIDAS: lessons learned from the first spaceborne atomic force microscope

M. S. Bentley[a], H. Arends[b], B. Butler[b], J. Gavira[b], H. Jeszenszky[a], T. Mannel[a,c], J. Romstedt[b], R. Schmied[a], K. Torkar[a]

[a] Space Research Institute of the Austrian Academy of Sciences, Schmiedlstraße 6, 8042 Graz, Austria
[b] European Space Research and Technology Centre, 2201 AZ, Noordwijk, The Netherlands
[c] University of Graz, Universitätsplatz 3, 8010 Graz, Austria

Corresponding author: M. S. Bentley (mark.bentley@oeaw.ac.at)

# Abstract

The Micro-Imaging Dust Analysis System (MIDAS) atomic force microscope (AFM) onboard the Rosetta orbiter was the first such instrument launched into space in 2004. Designed only a few years after the technique was invented, MIDAS is currently orbiting comet 67P Churyumov-Gerasimenko and producing the highest resolution 3D images of cometary dust ever made in situ. After more than a year of continuous operation much experience has been gained with this novel instrument. Coupled with operations of the Flight Spare and advances in terrestrial AFM a set of "lessons learned" has been produced, cumulating in recommendations for future spaceborne atomic force microscopes. The majority of the design could be reused as-is, or with incremental upgrades to include more modern components (e.g. the processor). Key additional recommendations are to incorporate an optical microscope to aid the search for particles and image registration, to include a variety of cantilevers (with different spring constants) and a variety of tip geometries.

# Keywords

# 1 Introduction

Sample return missions like Stardust [1], Genesis [2] and Hayabusa [3] represent the ideal method to study the micro and nanoscale properties of cosmic materials, however they are not always feasible for cost, engineering or programmatic reasons, or when *in situ* study of the material is required. In such cases instrument development focuses on replicating as much of the functionality as possibly within the mass, volume and power constraints of an *in situ* mission, whilst taking into account the need for considerable onboard autonomy.

Whilst several planetary missions have carried optical microscopes with resolutions down to a few micrometres per pixel (e.g. onboard Phoenix [4], Beagle-2 [5], MER [6] and MSL [7]), this is not sufficient to describe the sub-micron particle population or to study the surface morphology and roughness of particle surfaces. These properties can be extremely important to understand mineralogy, how light, gas and dust interact (particularly for comets), and how dust grains interact with each other (e.g. how adhesive they are, how readily heat can be transferred between them etc.). To achieve these goals a resolution of some nanometres is required. For a planetary mission, where limited sample preparation is possible and samples are likely to be non-conductive, an atomic force microscope (AFM) is the obvious candidate instrument to achieve this goal.

To date two atomic force microscopes have been launched into space. The Phoenix Mars lander [8] carried an instrument called MECA (the Microscopy, Electrochemistry, and Conductivity Analyzer) to Mars in 2007. MECA combined a wet-chemistry laboratory and an optical and atomic force microscope [4]. The MECA AFM successfully analysed samples of Martian soil, determining the size distribution [9] in conjunction with the optical microscope and observing a variety of particle morphologies.

The MIDAS AFM [10] on-board Rosetta [11] was the first AFM to be launched into space in 2004 and successfully made the first measurements during the commissioning phase shortly thereafter. However, Rosetta needed ten years to reach its target (Jupiter family comet 67P Churyumov-Gerasimenko). The goal of MIDAS is to collect cometary dust particles emitted from the cometary nucleus and to study their size, shape, morphology and related parameters with nanometre to micrometre resolution *in situ*.

# 2 MIDAS vs a regular AFM

Most atomic force microscopes operating in vacuum use the frequency modulated, rather than amplitude modulated, mode. This is because the high Q-factor of a cantilever in vacuum makes the feedback controller difficult to implement. Indeed the MECA AFM used this mode, monitoring the frequency change (caused by tip-sample interactions) of the cantilever via the corresponding phase shift [4]. A fast feedback loop was then used to maintain a constant cantilever height above the sample during scanning.

MIDAS, on the other hand, operates primarily as an amplitude modulated atomic force microscope (although contact mode is also available). MIDAS avoids the feedback issue by performing software-controlled point approaches to each pixel of the image. The result is a scanning system that can be slower than a traditional instrument, but that is more robust (feedback parameters do not need to be tuned for each sample). In addition, the approach (force-distance) curve can be recorded throughout an image, providing information about the physics of the tip-sample interaction, and thereby the physical properties of the sample.

This mode of operation, along with important parameters for operation, is summarised schematically in Figure 1. An image is created by a one-dimensional approach of the tip towards the substrate at each pixel position. The Z piezo is moved, under software control, until the cantilever/tip senses the sample via an amplitude change in dynamic mode, or a static deflection in contact mode. At this point the cantilever is moved backwards a preset amount (called the retraction distance, $z_r$) before an (unmonitored) X or Y movement is made to the next position. The value of $z_r$ is thus set relative to the lateral step, the scanner-sample slope and the expected sample height and topography. If the sample size is unknown, a cautious (i.e. large) value must be chosen to avoid striking the sample from the side during the lateral movement, as is shown at point 3 in Figure 1. As a result, most time during the image is spent with the tip far from the sample. A trade-off is therefore necessary between risk and image duration (and, ultimately, the science achievable, which is related to the number of images acquired during the mission lifetime!). For MIDAS a typical image scan takes several hours, while a MECA scan was performed in 30 minutes and modern terrestrial AFMs are able to scan at a rate of a few seconds per image.

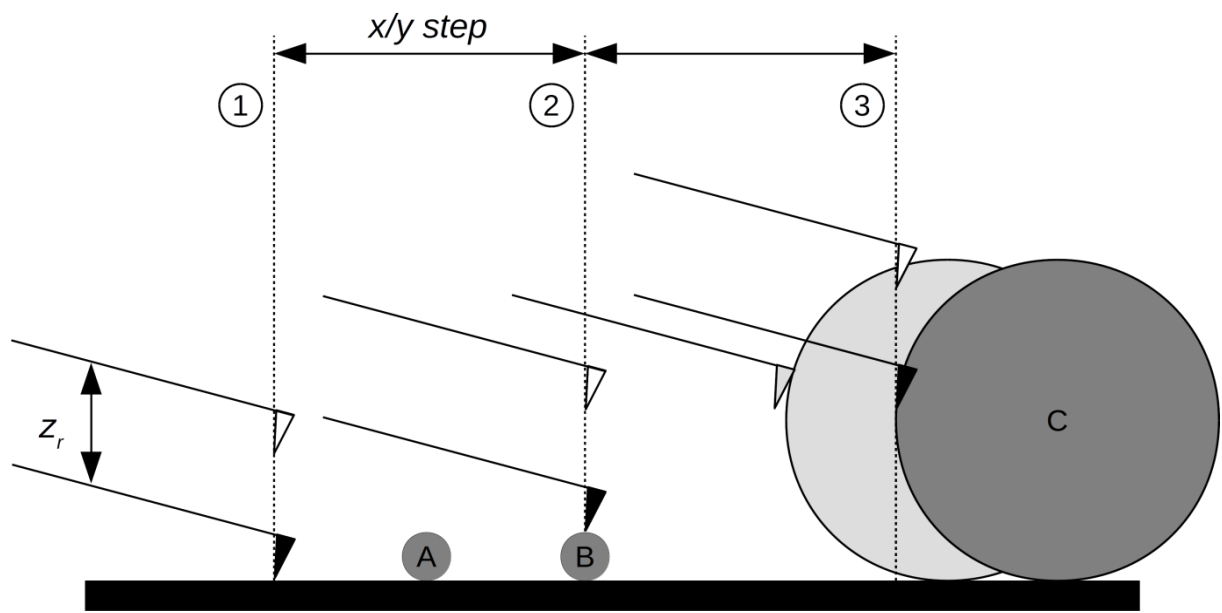

*Figure 1: A schematic of the MIDAS measurement technique. Each of the points 1, 2 and 3 represent point approaches to the substrate (bottom) at unique pixel positions in one row, separated by an X or Y step. After the surface is detected (black cantilever tip), the cantilever is moved away from the sample a distance $z_r$ (white tip). In the case of moving from position 1 to 2, particle A is missed, whereas particle B appears as a single pixel. In the case of particle C, the retraction height after position 2 is insufficient for a clear move to*

*position 3, and the large particle is hit from the side (grey tip), resulting in the particle being moved or distorted. On reaching position 3, MIDAS detects immediately that it is in contact with the sample and retracts from this point. The initial particle position is shown in light grey and the end position after unintended movement in dark grey.*

A further difference is that cantilevers are not replaced when the tips wear out, but that an entire array of cantilevers and tips is mounted to the scanner for redundancy, and the sample is positioned in front of the desired tip. This has some implications for the accuracy with which the same location on the target can be located with different cantilevers.

Nominal operations of MIDAS can be divided into two simplistic modes - exposure (in which a target is positioned in front of the dust funnel and the shutter opened), and scanning (in which the AFM is brought into contact with the sample and a line or image scan performed). Both of these require several mechanical operations to be completed. Collection targets are mounted on the circumference of the sample wheel, which rotates to move samples between the exposure and scanning position, and also provides coarse positioning in the Y direction (with respect to the AFM). The cantilevers are mounted on a linear array and the sample wheel can be moved laterally (along its axis) to select a cantilever and perform coarse X positioning. Scanning itself is performed by a high-resolution piezoelectric XYZ stage. MIDAS is thus a combination of AFM and a sample collection and handling system, with the added complication that, unlike a terrestrial instrument, it is often not known before a scan if there is a cometary sample present, and how large it is.

# 3 Lessons learned

MIDAS was designed only a few years after the initial development of the AFM [12]. Whilst developments in the design of, and theory underpinning, atomic force microscopy have progressed considerably in the meantime, the hardware design of MIDAS necessarily remained unchanged. Fortunately the flexibility afforded by software that can be changed in-flight means that new features have been added in the meantime, enabling modes of operation that were not originally foreseen. Both experience with the MIDAS Flight Model (FM) and Flight Spare (FS) instruments, coupled with the progress in terrestrial instrumentation, have led to a set of lessons learned which should be considered if a similar instrument is to be flown in the future.

## 3.1 Locating particles

A typical AFM can scan a rectangular field with a side length of some hundreds of nm up to 50 - 100 µm. In order to collect sufficient material, and to be physically manageable, MIDAS targets are much larger than this (1.4 x 2.4 mm). Therefore some method of locating particles is desired, since scanning the entire target would be a prohibitively time-consuming task. The MECA AFM on-board Phoenix addressed this by combining an optical microscope and AFM, and a well-defined sample delivery mechanism (the robotic arm and scoop). MIDAS has neither of these. An optical microscope was not possible within the instrument design constraints (mass, volume, complexity) thus AFM scans themselves have to be used to find particles.

Furthermore the dust collection rate and thus the number of cometary dust particles collected in a given period is unknown. The collection rate is a function of the local dust flux and size distribution at the location of the spacecraft, the dust velocity and the spacecraft pointing profile. A model based on the observations and data of Fulle *et al.* [13] was developed to predict this collection rate and the resulting target coverage and was used to plan exposures.

The results of this model for a one week dust collection shortly after Rosetta arrived at comet 67P (assuming continuous exposure) are shown in Figure 2. The number of particles per scan area has been calculated under various assumptions - this example is for low density particles in the upper limit given by [13], and an interpolation of the measured size distribution to smaller sizes. Spherical particles are assumed and placed randomly on the simulated target and an ideal, artificial, AFM image (assuming an infinitely sharp tip) is produced for a variety of sizes and resolutions. Table 1 lists the key parameters including the step size and retraction height. Of particular importance are the durations of scans of different resolutions. This process allows the scan parameters to be tuned in order to detect the maximum number of particles per unit time in a given scan and to preserve tips from damage by choosing an appropriate retraction height.

A second consideration arising from such a model is how much of the target needs to be scanned before a particle (of a given size) is found. In order to ensure unambiguously that a particle is of cometary origin a scan with similar resolution and coverage should be available both before and after exposure. Thus if the model predicts that a large scan area is required, a similarly sized area must be pre-scanned. Since scanning and exposing are mutually exclusive, and scanning with the AFM is the only way to find particles, the available time must be partitioned between pre-scans, exposures, coarse scans (to find particles) and follow-up scans (to image these particles at high resolution). Time is therefore one of the most precious commodities for MIDAS operations!

An independent method of determining if particles have been collected would have solved many of these problems. For example, even if an optical microscope could not resolve the smallest particles, placing constraints on the larger grains would already allow the microscope parameters (in particular the retraction height) to be tuned such that scans run much faster (without having to worry about hitting a large particle).

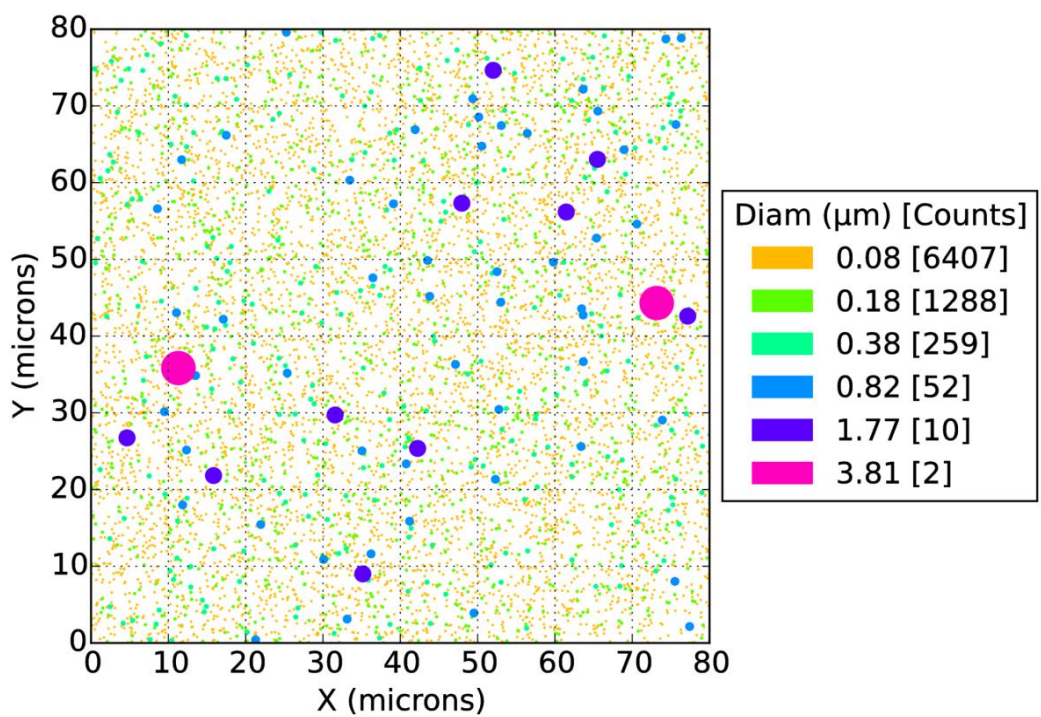

(a)

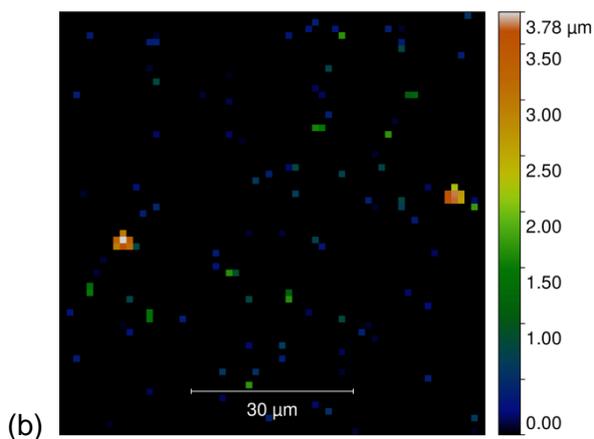

(b)

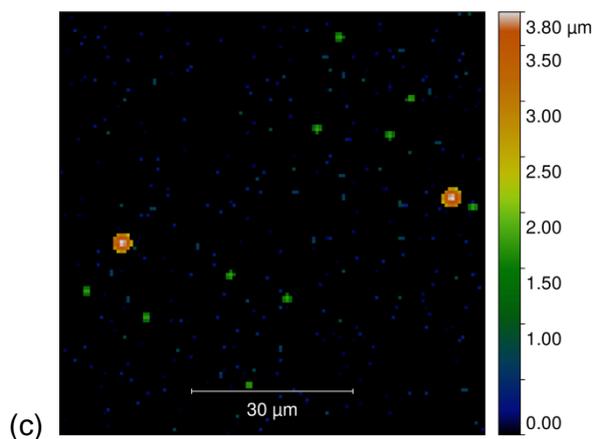

(c)

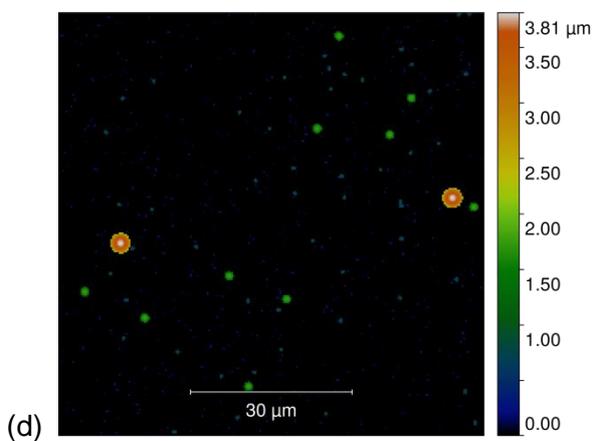

(d)

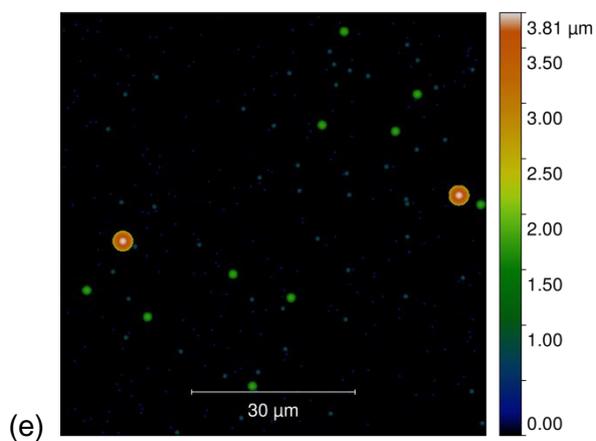

(e)

*Figure 2: An example output from the dust flux model, calculated for a hypothetical exposure between 2014-10-14 and 2014-10-21. The model accounts for spacecraft pointing and distance and the MIDAS funnel geometry. (a) The particles collected in each size bin are randomly placed onto the target plane. (b)-(e) Simulated AFM scans can then be generated from this distribution with different resolutions to evaluate how many particles would be imaged, and the corresponding scan time used to plan real operations. The scans shown here have resolutions of (a) 64x64, (b) 128x128, (c) 256x256 and (d) 512x512 pixels.*

| Image size (pixels) | X/Y step size (nm) | Retraction height (nm) | Duration (hh:mm:ss) |
| --- | --- | --- | --- |
| 64 x 64 | 1250.9 | 2501.8 | 4:42:38 |
| 128 x 128 | 625.5 | 1251.0 | 9:39:15 |
| 256 x 256 | 312.7 | 625.5 | 20:14:00 |
| 512 x 512 | 156.4 | 312.8 | 44:10:04 |

*Table 1: Key parameters corresponding to the simulated AFM scans shown in Figure 2 (b)-(e). In all cases the image size is 80 µm and the retraction height is taken as twice the X/Y step size.*

## 3.2 Image registration

One important issue when scanning small sections of a larger target is to ensure correct image registration. That is, to ensure that scans can be accurately located with respect to each other and that scans can be repeated. The position of a MIDAS tip with respect to an absolute position on the target is given in X by the lateral position of the wheel (and linear stage on which it is mounted) and the X coordinate of the XYZ stage. Similarly the Y position is given by the segment (1-1024) to which the wheel is rotated (determined by a differential encoder) and the Y coordinate of the stage. An additional step is needed when comparing images made with different tips; the relative offset between any two tips must then be taken into account.

Various strategies were considered to achieve this during the MIDAS design phase, for example micro-machining a unique pattern onto the substrate which could be read out during imaging, but a scale-independent solution was not easy to implement. Part of the problem is simply that the biggest issues with image registration occur for smaller images (e.g. less than a few microns in size) and it is hard to imagine a pattern that could be readily engraved onto the substrate and yet be useful at a variety of image scales.

A related issue is that the MIDAS wheel encoder has 1024 positions, giving a 0.35 degree angular resolution and 13 positions over the 2.4 mm length of the target. Unfortunately the resulting accessible area shifts with each wheel movement by a distance greater than the largest scan field (~100 µm). As a result, not all of any given target is accessible to the microscope as shown in Figure 3. In principle this should not be a problem since the expected particle size distribution should lead to a large number of smaller particles that can be found everywhere on the target; even with this limited access the target is in any case much larger than the area that can be scanned in a reasonable amount of time. However, in at least some exposures larger particles were collected that appear to have fragmented shortly before, or on, impact with the target. In these cases it would have been ideal if the entirety of the fragment collection could have been imaged in order to reconstruct, or at least constrain, the properties of the parent particle. An additional consideration is that the largest particles collected by MIDAS to date are some tens of microns in extent and inevitably some lie close to the edge of the scan field in the Y direction, and thus only part of the particle can be imaged.

At the time of the instrument proposal the encoder was state-of-the-art, but it is probable that an alternative product would be readily available twenty years later! In addition, as already mentioned in the context of locating particles, an optical microscope would have also been useful to confirm the precise location of each image with respect to the other, and a system of alignment patterns visible to such a microscope would have been much simpler to produce than an equivalent pattern for the AFM.

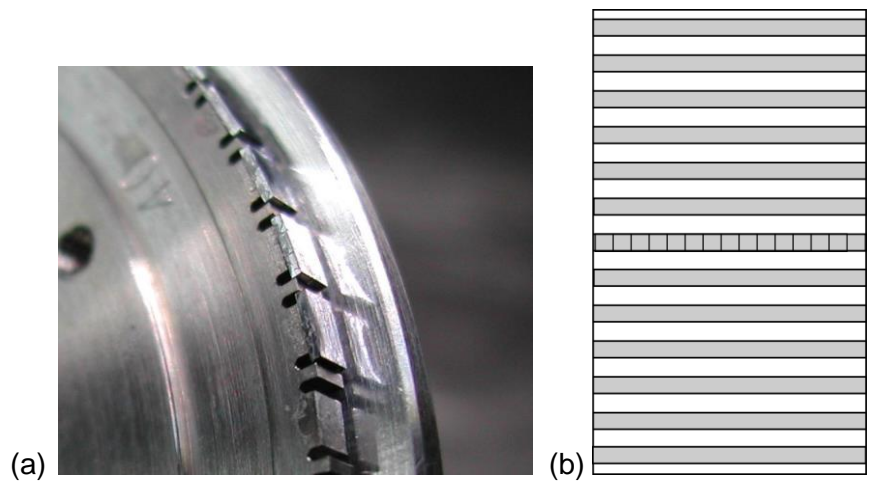

*Figure 3: (a) A ground model of the MIDAS wheel showing collection targets mounted on the perimeter. This wheel rotates with a finite step size to position the target. (b) A diagram of the accessible target area (shaded "stripes") due to these finite steps. The small squares on the central stripe show the largest area that a single scan can cover.*

### 3.3 Exposure and scanning

As already mentioned, MIDAS cannot expose one target and scan another arbitrary target simultaneously. This is a result of the basic design, which arranges targets on the circumference of a wheel. In addition, the original calculations performed when MIDAS was designed (based on a different comet and more time spent close to the nucleus) resulted in exposure times on the order of hours and days, not weeks.

In the current design the 16 cantilevers are symmetrically centred on the funnel with a spacing of 1.6 mm (slightly larger than the target width of 1.4 mm) such that the wheel is positioned directly in between the two central cantilevers in the exposure position. This limits the possibilities of simultaneous scanning and exposure to the two central cantilevers, and then when the target is only partially exposed. Due to the low dust flux seen by MIDAS, this mode of operation, with a reduced exposure area, has not been used to date.

One minor change that could be considered for an iteration of this design would be to offset the cantilever array such that in the exposure position one cantilever is already centred on the sample wheel (or of course to use an odd number of cantilevers and keep the symmetric design).

### 3.4 Spacecraft charging

A power-law size distribution is typically assumed for cometary dust, with a more shallow distribution for the sub-micron particles that are of primary interest for MIDAS expected from *in situ* data from comet Halley [14]. Thus for every large particle scanned, a myriad of smaller particles should be expected. Early measurements with MIDAS at the comet were quite the opposite - few micron-sized particles were seen. The dust funnel has a field-of-view of thirty degrees and spacecraft off-pointing was typically much less than this, and so pointing is not likely to have been responsible. The spacecraft velocity was also rather low during this phase and aberration effects should also have not limited dust collection.

Data from other instruments on-board Rosetta were of course used to constrain the possible dust collection rate - for example GIADA [15] detects single (large) particle impacts and also measures the cumulative deposition of smaller grains, whilst the COSIMA dust mass spectrometer [16] employs a similar expose/analyse strategy to MIDAS and can image targets with an optical microscope with a resolution of 14 microns. Whilst on balance the various instruments confirmed a more shallow size distribution for smaller particles, it soon became clear that there was a strong size and/or instrument selection effect.

The most plausible explanation for this appears to be that the local plasma environment of the spacecraft resulted in a negative spacecraft (and, by analogy, dust) potential [17]. Such a potential results in strong electric fields that can either stop or deflect small particles from entering the instrument. It has also been suggested that this charge effect can fragment larger dust particles, resulting in slow-moving showers of such fragments being detected at the spacecraft [18]. A review of the Rosetta spacecraft geometry also revealed that the MIDAS funnel is positioned on the nadir platform rather close to a medium gain antenna and a sun acquisition sensor. Both of these extend from the spacecraft wall further than the MIDAS funnel and have sharp edges, and thus could have contributed to charged grain deflection. A detailed study of this effect is pending.

### 3.5 Planning and intent

Most instruments onboard the Rosetta orbiter have observations planned far in advance (since they require pointing that is fixed in the medium term planning process, some weeks or months before the observation itself) or very limited planning (for monitoring instruments). But some instruments, in particular COSIMA and MIDAS, are an interesting combination of both. During periods where other instruments are actively observing (e.g. during a close flyby of the nucleus), dust collection instruments are exposing. Immediately following this, the exposed targets must be imaged and follow-up measurements commanded. The resulting planning strategy is more akin to a lander mission than a typical mission, since it is more exploratory and received data need to be folded back into future planning on an as-short-as-possible time scale.

Due to the planning horizons (typically several weeks), at the time a follow-up measurement was commanded, several other planning sequences were already scheduled. Since MIDAS operations are very dependent on the given cantilever and its history (each with a unique resonance frequency and tip properties) there was a risk that these intermediate sequences could damage (wear or contaminate) the tip. Unlike the MECA AFM, in which only a single cantilever could be used (until blunt, after which it was physically removed), MIDAS can utilise any cantilever at any time. A strategy was therefore adopted in which up to three cantilevers were operated, one per planning cycle, such that when planning follow-up operations the chosen cantilever was never scheduled to be used in the intervening periods.

Because of this short-term planning, with decisions often made as late as possible (with respect to planning deadlines), it is easy for planning intent to become opaque to anyone not directly involved in the process. This has implications for data archiving, where in an ideal world the entire planning process should also be recorded. An attempt to solve this was introduced in the MIDAS team by documenting weekly planning using IPython (now Jupyter) Notebooks, which combined the necessary code-based analysis and command generation with textual description of the rationale behind a given observations. These notebooks will then either be deposited in the ESA Planetary Science Archive, or made available in a public online repository (e.g. GitHub).

## 3.6   Autonomy and robustness

A typical AFM is a rather interactive instrument and action is often needed if the tip loses contact with the surface, if the feedback parameters are not set correctly, or simply if the image appears distorted or otherwise incorrect. For a spaceborne instrument this is not possible due either to the large one-way-light time, long planning horizons, or both. Rosetta awoke from deep space hibernation at a distance of 5.4 AU, giving a round-trip signal time of 90 minutes. Limited power and data at this distance meant that payload usage had to be carefully planned. As such, instrument commanding had to be completed and verified several weeks ahead of execution, with additional operations to be executed in the meantime.

As a result, exposure and scanning operations were divided into distinct blocks of telecommand sequences which had to be independent of each other and internally robust. This meant that if one block failed, the subsequent blocks should not be affected. The resulting sequence is often not optimal, but the gain in robustness is essential; for example some mechanisms are actuated when they *should* already be in position, but this is not guaranteed if previous operations have failed.

In addition to robust ground-based commanding, some degree of autonomy was also required to achieve the highest resolution scans. The coarse positioning mechanisms used by MIDAS are accurate to within some tens of micrometres, but a nanometre resolution scan of a micron-sized particle cannot easily be achieved by direct commanding. To overcome this, a dedicated feature recognition algorithm was implemented. This uses the topographic data from a coarse scan to identify features matching certain criteria and perform subsequent higher resolution imaging.

Detailed features in AFM data are often not immediately visible due to large scanner-sample slopes and thermal distortions. To remove such artefacts the MIDAS on-board software (OBSW) first performs a least-squares plane subtraction and (optionally) median line subtraction to the topographic data. A commandable threshold is used to identify possible features and a number of optional criteria used to rank and select a feature for a subsequent scan. These include the number of pixels making up the feature, its average height and aspect ratio. Figure 4 shows an example of these processing steps applied to an image from the MIDAS flight model. This algorithm was originally designed to return statistical information on the particles collected on a target in case insufficient data volume was available to return complete images. In reality the high data volume available has meant that this function was not needed. Instead it is typically used by commanding a medium resolution image of a previously identified particle (of which a high resolution zoom is desired) and using this feature recognition capability to autonomously locate the feature precisely within the frame of the scan and set the origin and step size for a subsequent zoom. This is done without actuating any of the coarse mechanisms of MIDAS and hence the accuracy is very high.

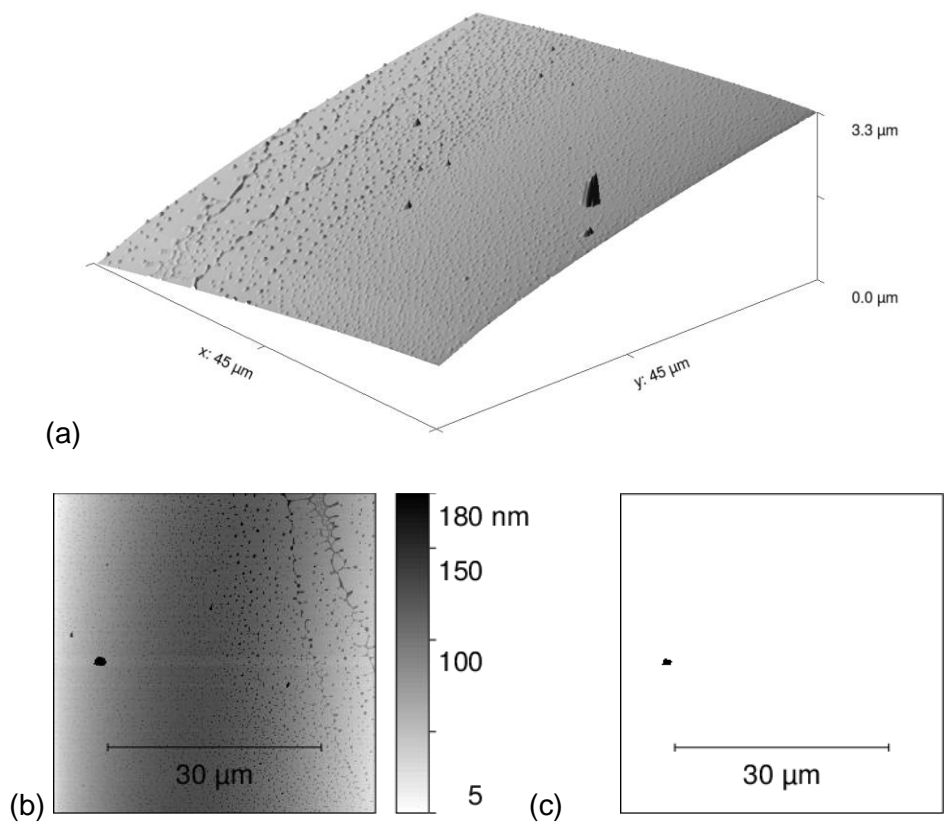

Figure 4: An example of an image from the Flight Model and the steps taken by the MIDAS OBSW to identify and perform a zoom scan of the dust grain. 4(a) shows a 3D view of the scan with 5 times vertical exaggeration - the slope with respect to the scanner is evident, as

*is some thermal distortion. 4(b) shows this image after plane subtraction and median line correction. A threshold of 50% of the maximum height is applied to give 4(c).*

## 3.7 Vibration

An atomic force microscope is rather sensitive to various external factors, in particular temperature and vibrations. This is why many commercial terrestrial instruments are mounted on damped vibration tables and shielded within enclosures. As well as operating during various mechanism actuations onboard the Rosetta orbiter, MIDAS first had to survive the launch loads. The instrument thus incorporated various locking mechanisms to ensure that the sensitive components would not be damaged during launch. As discussed in [10] these functioned extremely well, however one capacitive sensor was found to give erroneous readings after launch, meaning that closed loop control of the scanner head was possible only in the Y axis. A "hybrid" mode was then devised where scans were made primarily with Y being the fast direction and X was used in open loop mode.

Even after launch a spacecraft has a variety of mechanisms that can induce unwanted vibrations. For example thruster firings, the high gain antenna and solar array drive motors, reaction wheels and various instrument subsystems (including the Stirling cycle cooler of the visible and near infrared spectrometer VIRTIS [19] and the robotic target manipulation system of COSIMA [16]). To reduce the effect of these vibrations on MIDAS scans, the entire microscope was mounted on four silicone dampers. This damping system was locked during launch to avoid damage and unclamped shortly after Rosetta reached orbit. Theoretical predictions of the damping efficiency suggested a damping factor (at frequencies above a few tens of Hertz) of approximately one order of magnitude, with a main resonance close to 5 Hz. No vibrations onboard Rosetta were expected at this frequency and the damping becomes efficient at frequencies above a few times this value.

The effectiveness of these dampers was investigated both before and after launch by use of a dedicated vibration monitoring mode. In this mode a cantilever is brought into contact with a sample in the static mode in which the cantilever is not oscillated but its deflection monitored. The amplitude is then logged at as high a frequency as the data processing unit (DPU) allows. The result is a frequency spectrum of any external disturbances.

The performance of the MIDAS Qualification Model was evaluated during a dedicated vibration test in which sinusoidal and random vibration inputs were used on all axes (0.1 to 20 milligee RMS, 1 to 400 Hz). These tests were performed with the vibration system locked in its launch configuration and then again unlocked to evaluate the effectiveness of the damping system. In order to perform the unlocked tests under Earth gravity, the weight of the microscope stage had to be partially offloaded by supporting it on a long, flexible, wire.

Example plots are shown in Figure 5 where the change in amplitude scale after unlocking the damping system should be noted. The damping factor predicted by theory was indeed confirmed by these tests. In conclusion, the damping system performed according to expectations, with residual disturbances showing low level white noise corresponding to 0.02 digits at frequencies >50 Hz. This scales to ~2 digits or ~0.2 nm at the specified input level, well below the threshold to cause severe interference with AFM imaging.

A similar test was performed on the Flight Model after launch during a dedicated interference session in which sets of instruments whose operation could potentially interfere with each other were activated and monitored. For MIDAS this meant operating in conjunction with the VIRTIS cryo-coolers, the OSIRIS filter wheel and the COSIMA target manipulation unit. A similar series of vibration monitoring tests was performed and no major interference source was identified. Nonetheless in early comet operations MIDAS scans were scheduled to avoid thruster firings, including reaction wheel off-loadings. In later phases of the mission scans were performed during these operations and demonstrated that no effect was visible in the image data, and so only larger firings (during orbital correction manoeuvres) were avoided. This considerably simplified operations and allowed for longer continuous scans.

(a)

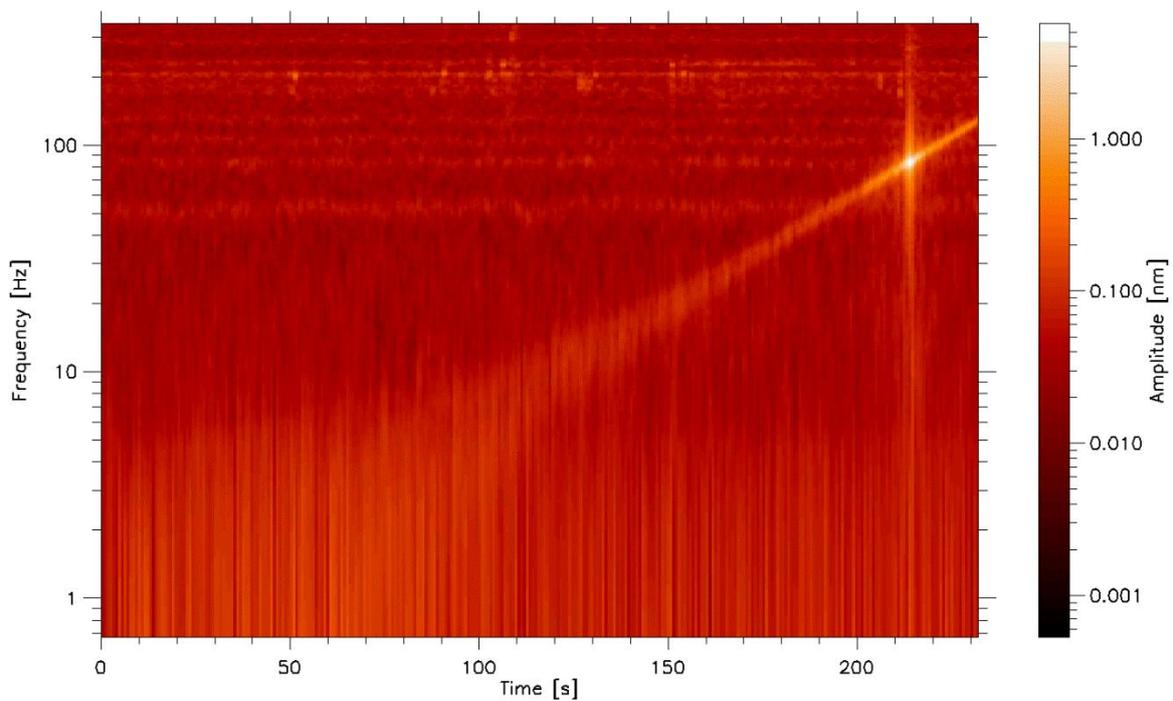

(b)

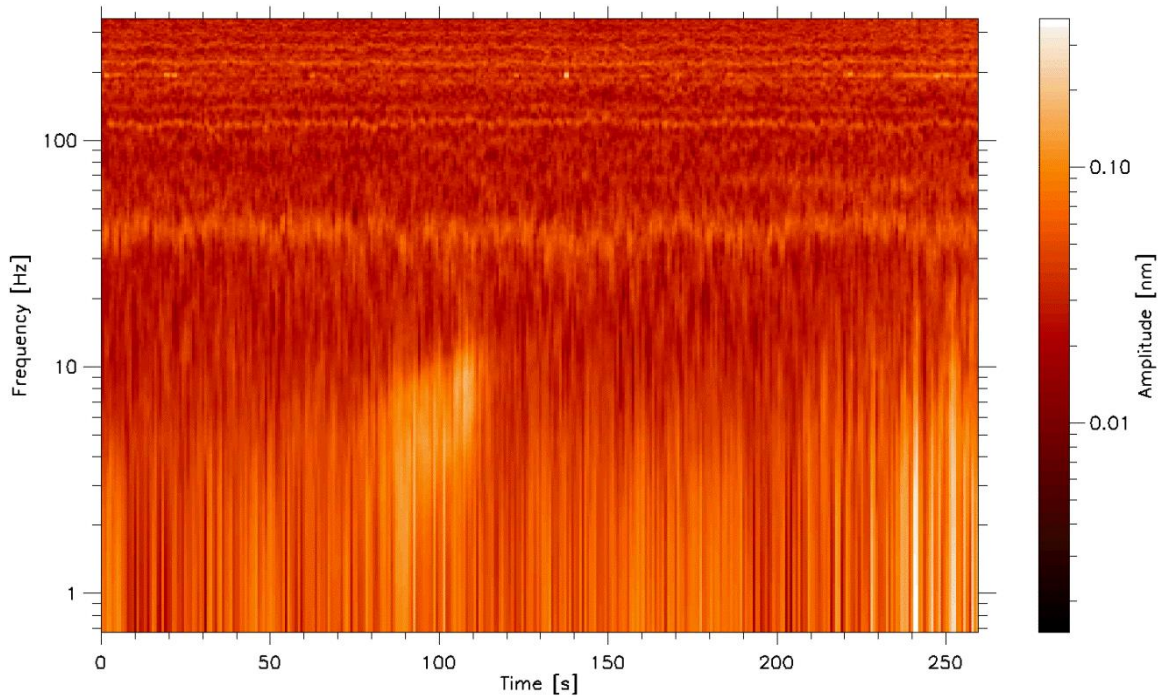

*Figure 5: Frequency spectra made during ground-based vibration testing of the Qualification Model. (a) Shows the results for the vibration damping system in its locked (launch) configuration and (b) shows the same after unlocking. Note the difference in amplitude scale.*

## 3.8  Piezo distortions

The heart of an AFM is the XYZ stage that moves the tip with high precision in three dimensions. This is usually built around a set of piezoelectric elements which are both reliable and precise, but are susceptible to both temperature and non-linearity effects.

Temperature drifts are often evident in atomic force microscopy. Such drifts can appear in the XY plane (as evidenced by scans of regular calibration targets) or as a curvature of the plane in the Z direction (in particular in the "slow" direction). Although the spacecraft guarantees a specific temperature range for the Temperature Reference Point, variations within this range can occur on short timescales due, for example, to spacecraft pointing or simply restarting the microscope. MIDAS does not have active thermal control, but the design is such that temperature variations are minimised. Such variations are, however, evident in most scans, especially when the spacecraft is close to the Sun and the pointing is changing such that the MIDAS funnel becomes more or less illuminated. In many cases height variations can be removed by careful polynomial background subtraction, or even simple plane subtraction. Lateral distortions are, however, harder to identify and remove on regular substrates, which do not have any well-defined features.

There are several reference sensors on-board MIDAS which report subsystem temperatures in housekeeping telemetry. Unfortunately none of these are positioned very close to the cantilevers, where the effect of temperature variations is likely to be most significant. The MECA AFM included a reference temperature sensor mounted on a short cantilever; such a sensor should definitely be included on any future space-borne AFM. The recently developed technique of Scanning Thermal Microscopy, which uses a nanofabricated thermal probe with a resistive element, could also be implemented. This not only allows the thermal properties of the sample to be measured, but also the temperature variations directly at the tip.

The MIDAS scanner-head was designed to operate in a closed loop configuration, with a capacitive sensor system used to ensure that any non-linearities in the piezo behaviour, or influences of temperature drift, were removed. However one of the two channels of this system was found to be unresponsive after launch. This limited operation to purely open loop, or else one channel could still be operated in closed loop mode (the "hybrid" mode).

Due to the various influences on the piezo movements, calibration standards are mounted on MIDAS. Whilst primarily there to ensure accuracy calibration into physical units, they can also be used for distortion correction. The key standard is a rectangular grid of 1.1 x 1.1 µm² squares, which are 3 µm apart and have a height of 900 nm. Thus the lateral distortions, which partly come from temperature variations but also from piezo nonlinearities and other effects, can be determined by scanning the well-known target (Figure 6 (a)). With these scans the positions of the calibration target squares can be corrected to be vertically and horizontally aligned. This not only minimises the influence of thermal variations, but also the piezo nonlinearities (Figure 6 (b)). The most distorted part of the image occurs in the first few lines (top) and at the start of each new line (left). A good reconstruction is not possible in these areas and the resulting image is cropped to remove them. By regular scanning of this target, and monitoring the existing temperatures, these corrections can be applied also to scientific images. The use of such calibration samples is thus essential and additional targets could be considered covering a range of sizes and resolutions.

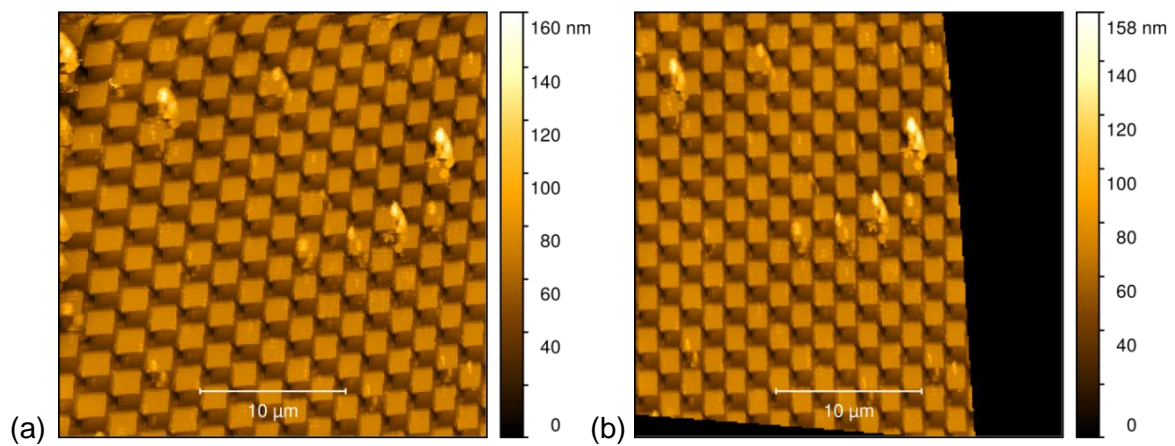

*Figure 6: (a) An uncorrected hybrid mode 30x30 µm² topographic image showing distorted calibration structures and (b) a distortion-corrected image showing rectangular shaped regular structures. These images were acquired during tests on the Flight Spare, but a similar procedure is used for flight data calibration.*

### 3.9 Cantilever and tip design

The piezoresistive cantilevers and tips designed for MIDAS were state-of-the-art developments when the instrument was constructed, but (inevitably) twenty years of commercial experience on the ground has produced significant advances. In particular the range of tips available has dramatically increased, both in terms of materials, shape and sharpness. For example, a variety of tip shapes are now available, some with a high aspect ratio that would be very suitable for imaging the rather rough "fluffy" particles that are typical for cometary dust. In general equipping an AFM designed to study unknown samples with a range of tip geometries would be useful.

Another development that has taken place since the initial design of MIDAS is that the theory underpinning dynamic AFM has advanced considerably (see e.g. [20]). This allows for quantitative, or semi-quantitative, understanding of the various tip-sample interactions. To interpret such measurements, a good knowledge of the cantilever spring constant is always a prerequisite. This can be estimated for the MIDAS cantilevers via the Sader method, by using the dimensions and measured resonance frequency [21]. Indeed a finite element model of the cantilevers, taking into account the detailed shape, results in an almost identical value of approximately 300 N/m. This is rather stiff and is not optimal for all measurements that could be possible with the MIDAS hardware (for example force spectroscopy); some of these modes were not even invented when the instrument was developed and could not have been foreseen. Nonetheless a future AFM should carry a variety of spring constants in order to probe a wide range of materials in different operating modes.

## 4  Outlook

MIDAS, the first spaceborne atomic force microscope, has performed remarkably well in a rather challenging environment. Even considering the twenty years in between design and operation at the comet, software upgrades and modelling have allowed more operating modes to be defined than originally planned and additional data to be extracted. The full data set should reveal new secrets of cometary dust at the nanometre scale for decades to come.

Based on the operational experience with MIDAS and advances in terrestrial atomic force microscopy, there are several enhancements that could be envisaged for a future spaceborne AFM. These fall into two categories - minor modifications to the existing design that would allow a re-flight without major additional qualification, and more substantial changes. These are summarised in the following paragraphs.

Minor modifications

1. Include a variety of cantilever spring constants - this helps probe a range of materials and allows additional operating modes, such as force spectroscopy, without any changes to the hardware design.
2. Include a variety of tips - including very sharp, high aspect tips that may be more fragile but that can best image samples with high topography and topographic gradients.

3. Incorporate a modern processor - this would dramatically speed up the scanning process; a multi-core CPU would allow a dedicated core for scanning, without having to interrupt the process for handling of telemetry and telecommand requests etc.
4. Position the cantilever array such that one cantilever is aligned with the funnel - as a result this single cantilever could scan the opposite target to that being exposed. This would allow more time for scientific scans by enabling careful pre-scans to be performed during exposures.
5. Include a modern wheel encoder with finer steps, such that the sample wheel could be rotated with a single step smaller than the maximum image size (~100 μm). This would allow for continuous coverage of each target.
6. Include a reference temperature sensor mounted on the cantilever array, to monitor the real temperature of the cantilevers for on or off-line compensation.

Major modification

1. Include an optical microscope to locate (and potentially avoid) large particles, to target medium sized particles and to assist in image registration. In fact this need not be such a major modification since the current design has space on "top" of the instrument where a camera and illumination source could be conveniently located. However the best solution would integrate the microscope with the AFM scanner head such that optical and atomic force microscopy images are completely co-registered. This would require a more substantial design revision.

In conclusion in the twenty years since its design, terrestrial AFM and cantilever/tip fabrication techniques have come a long way and some of these updates could simply be plugged into the MIDAS design to have a much more modern instrument. Atomic force microscopy should certainly be considered for any future mission where the characterisation of materials at the nanometre scale is important, and lessons learned from both MIDAS and MECA should be incorporated in any future development.

# 6 Acknowledgements

Rosetta is an ESA mission with contributions from its member states and NASA. We also thank the Rosetta Science Ground Segment and Mission Operations Centre for their support in acquiring the presented data. All data from MIDAS will be made available through the ESA Planetary Science Archive (PSA).

T. Mannel gratefully acknowledges the Steiermärkische Sparkasse and University of Graz for financial support. R. Schmied acknowledges the Austrian Research Promotion Agency (FFG) for financial support under the ASAP-10 programme (project number 844393).

MIDAS became possible through generous support from funding agencies including the European Space Agency PRODEX programme, the Austrian Space Agency, the Austrian Academy of Sciences, ESA/ESTEC, and the German funding agency DARA (later DLR).

We also thank two anonymous referees for their comments and suggestions which greatly improved the readability of this manuscript.

# 7 Vitae

**Thurid Mannel** is a PhD student of the University of Graz working in the MIDAS team at the Space Research Institute of the Austrian Academy of Sciences. Her main research area is the analysis of cometary dust, in particular with respect to the morphology at the nanometre scale. She received her M.Sc. in Physics from the Technical University Munich.

**Klaus Torkar** received the Dipl.-Ing. and Dr.Techn. degrees from Graz Technical University (TUG), Graz, in 1975 and 1978. Since 1975 Staff Member and since his retirement in 2013 Guest Researcher with the Austrian Academy of Sciences, Graz. Professor with TUG since 2005. He was Principal Investigator (PI) of several ion emitter experiments for spacecraft potential control and for the cometary dust scanning force microscope MIDAS/Rosetta (2011-2013), Co-PI for the ion camera PICAM on Bepicolombo. He is (co-)author of 133 refereed scientific publications. Interests include spacecraft potential experiments, ion-chemical modeling, spacecraft–plasma interaction, and investigations of cometary dust.

**Mark Bentley** is a planetary scientist at the Space Research Institute of the Austrian Academy of Sciences. His research investigates surface processes on Solar System bodies via laboratory experiments, modelling and space mission data. Most recently he has been Principal Investigator of the MIDAS atomic force microscope onboard Rosetta. Mark has an MPhys in Physics with Space Science and Technology from Leicester University and a PhD in planetary science from the Open University, UK.

**Harald Jeszenszky** is an engineer at the Space Research Institute of the Austrian Academy of Sciences. As well as being technical manager of MIDAS and project managed for the MMS/ASPOC instrument, Harald is responsible for software development and hardware integration and testing for a variety of space instruments.

**Roland Schmied** studied technical physics on the Graz University of Technology and received his MSc and the PhD on the Institute for Electron Microscopy and Nanoanalysis working now as a PostDoc at the Austrian Academy of Sciences on the modeling and calibration of MIDAS atomic force microscope. His main research area is the materials characterization at the nanometre scale and its sample preparation and the morphological analysis of cometary dust.

**Jens Romstedt** joined the European Space Agency (ESA) in 1996. Since then he has been involved in prototype and flight instrument development for various space missions. In the last few years he has participated in the evaluation and study of future planetary missions within various ESA programmes. In addition he has been the technical lead for the development of critical technologies for robotic exploration of planetary surfaces.

**José Manuel Gavira Izquierdo** currently leads the Mechatronics and Optics Division, in the European space agency (ESA) that oversees and coordinates R&D in the fields of Automation and Robotics, Optics/Optoelectronic and Life/Physical-sciences for all ESA applications in Earth Observations, Science and Robotics Exploration, and Human Spaceflight Programmes. He hold a Master degree in Space System Engineering by Delft university and a BSc in Mechanical Engineering by the Politecnic Univ of Madrid. Mr Gavira has over 30 years' experience in space activities, first as Mechanical engineer in industry to develop structures and Mechanism for space missions.

**Herman Arends** has a BSc in mechanical engineering and worked at ESA/ESTEC for over thirty years, primarily on the design and test of plasma sources and probes for electric field measurement. In particular he worked on and had the responsibility of the mechanical design, assembly and environment tests of ion- and electron emitters for several space missions, and of the MIDAS atomic force microscope on-board Rosetta.

Passport-style photos:

| | | |
|---|---|---|
| 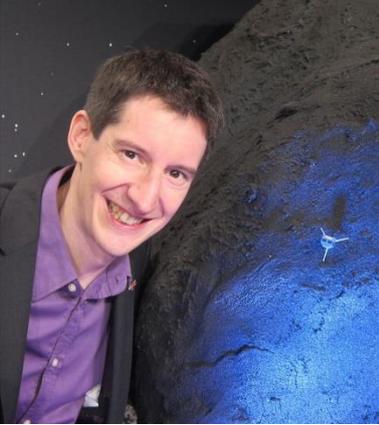 | 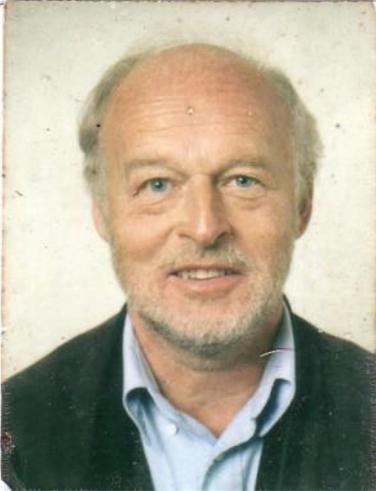 | 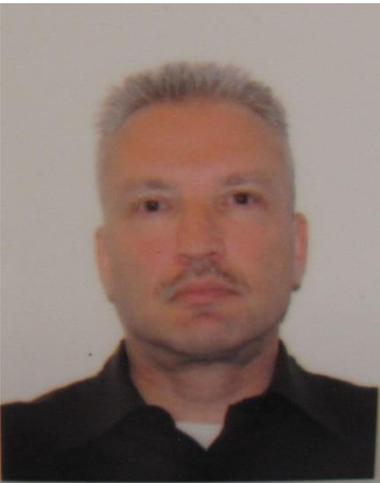 |
| M. S. Bentley | H. Arends | B. Butler |
| 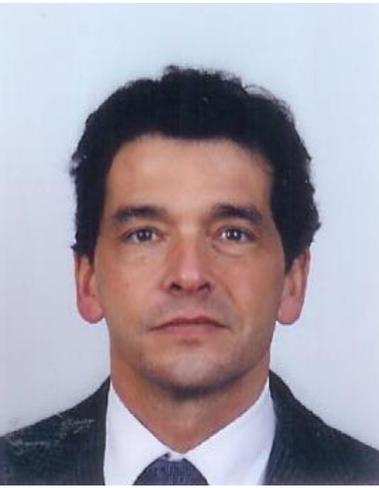 | 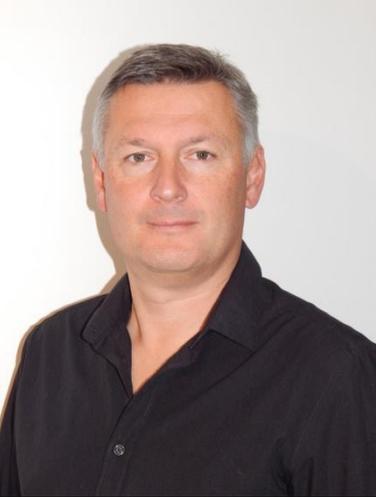 | 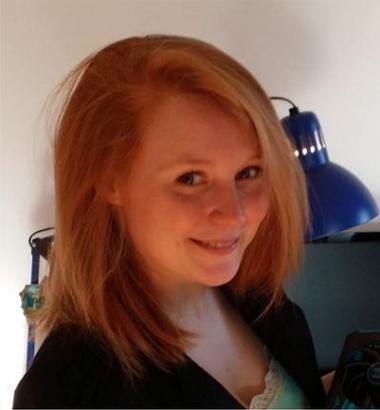 |
| J. Gavira | H. Jeszenszky | T. Mannel |
| 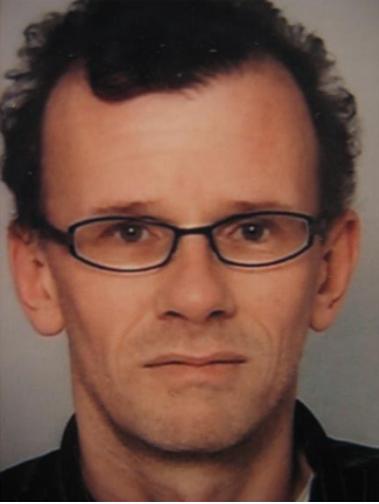 | 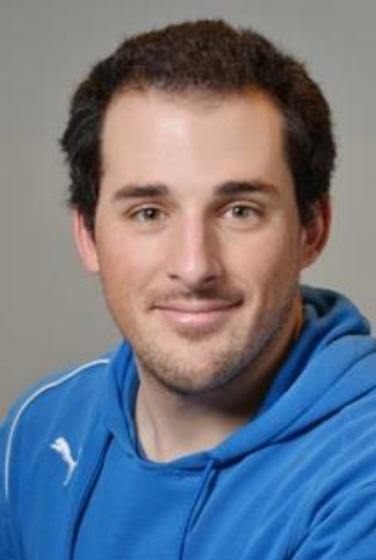 | 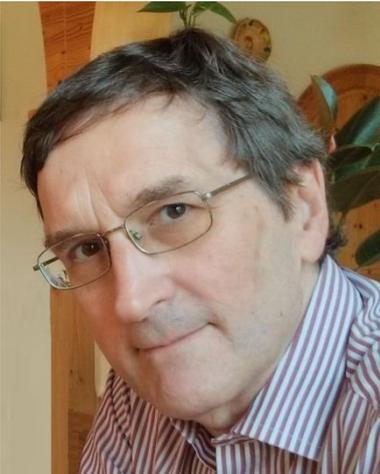 |
| J. Romstedt | R. Schmied | K. Torkar |